# Cell pelotons: a model of early evolutionary cell sorting, with application to slime mold *D. discoideum*

Hugh Trenchard
Independent researcher
805 647 Michigan Street
Victoria BC Canada V8V 1S9
h.a.trenchard@gmail.com
250-472-0718
ORCID ID 0000-0003-3784-4349

**Abstract**
A theoretical model is presented for early evolutionary cell sorting within cellular aggregates. The model involves an energy-saving mechanism and principles of collective self-organization analogous to those observed in bicycle pelotons (groups of cyclists). The theoretical framework is applied to slime-mold slugs (*Dictyostelium discoideum*) and incorporated into a computer simulation which demonstrates principally the sorting of cells between the anterior and posterior slug regions. The simulation relies on an existing simulation of bicycle peloton dynamics which is modified to incorporate a limited range of cell metabolic capacities among heterogeneous cells, along with a tunable energy-expenditure parameter, referred to as an "output-level" or "starvation-level" to reflect diminishing energetic supply, proto-cellular dynamics are modelled for three output phases: "active", "suffering", and "dying or dead."  Adjusting the starvation parameter causes cell differentiation and sorting into sub-groups within the cellular aggregate. Tuning of the starvation parameter demonstrates how weak or expired cells shuffle backward within the cellular aggregate.

**Keywords:** early evolutionary stage; energy-saving mechanism; heterogeneous metabolic outputs; output phase; peloton; slime mold; *Dictyostelium discoideum*
**MSC 92-08**

**Highlights**

- A non-chemotactic model of cell-migration and sorting is presented
- Model includes three main parameters:
    1. Cell motile speed or equivalent metabolic capacity, or diminishing energetic supply
    2. Range of individual cell maximum sustainable outputs
    3. An energy-saving mechanism
- Model demonstrates three main phases of collective behavior
- Model suggests an early-evolutionary stage for cellular slime-molds

## 1. Introduction

Collective cell migration is widely understood to involve extra-cellular cues, whether provided by food, other chemical gradients, mechanical forces or signal relays; the cues are sensed by migrating cells and translated into directed motion (Forgacs and Newman, 2005, p. 157; Weijer 2009). All migrating cells can sense and respond to guidance signals; alternatively, certain leader cells can sense the signals and then induce others to follow via chemical signals or physical, inter-cellular interactions. Within the last decade or so, collective cell migration has been explained through non-chemotactic mechanisms, such as durotaxis, in which cells alter their directed motion according to gradients in the stiffness of extracellular matrices (Sunyer et al., 2016).

Generally, however, non-chemotactic mechanical mechanisms are currently not well-understood (Rosa-Cusachs et al., 2013). Here we consider a novel non-chemotactic primordial mechanism involved in driving proto-cellular migration, proposed to have emerged at an earlier evolutionary stage than did chemotactic processes. This suggestion naturally invites the question: what exactly were the mechanical processes that facilitated proto-cell attraction, aggregation, and the emergence of coordinated and directed collective motion?

In response to this question, a model is proposed of an earlier evolutionary process in which proto-cellular migration is driven by mechanical means, rather than chemical attractors. The model is proposed to be applicable to an early evolutionary stage common to many biological systems. Here we examine the model in the context of the aggregate (slug) stage of the slime mold, *Dictyostelium discoideum*.

In this model, an energy-saving region extends in a wake behind and to the sides of each individual cell. This 'wake region' engages an energy-saving attractive process that facilitates cell-coupling because it is less energetically costly to occupy those regions. The theory that informs this model suggests an attractive process in which proto-cells originally occupied, by random motion, optimal wake-region positions. Then, due to fitness ~~adaptive~~ advantages conferred by occupying these low-energy regions, cells evolved further mechanical and chemotactic mechanisms enabling them to find optimal positions.

A useful analogue to this energy-saving attractive process is observed in non-organic systems involving collections of particles under the influence of gravity and water turbulence. Here, particles following in the path of others attract to those ahead (i.e. 'couple') by accelerating into a drafting zone. The process is known as "drafting, kissing, and tumbling" (Wang et al., 2014), as shown in Fig. 1.

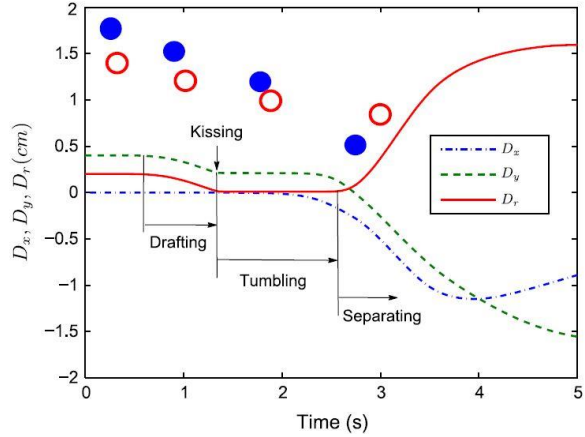

**Fig. 1.** Separation history of two particles falling in a viscous fluid. A following particle (blue) accelerates in a low-pressure region (drafting) behind a leading particle (red), makes contact ("kisses"), tumbles and separates. Ordinate-axis indicates distance between particle surfaces ($D_r$) in terms of $D_x$ (transverse) and ($D_y$) (longitudinal) coordinates. (Wang et al., 2014, Fig 5; reproduced with permission).

In this process, particles may temporarily align vertically atop each other, but horizontal alignments are more stable than vertical ones; particles can also rise upward into wakes (drafting regions) due to turbulence and can cluster in aggregations that are spatially separated (Fortes et al., 1987). Particles of varied sizes can also be involved in drafting dynamics (Wang et al., 2014). The shared dynamics of these non-organic particles and the organic proto-cells under consideration in this study suggests that their fundamental principles of motion are common. Thus, it is possible to state that energy-saving mechanisms that facilitate particle attraction and coupling operate similarly for cellular migration.

Moreover, since the dynamics of both inorganic particles and organic cells can involve attraction and coupling, it is proposed that an energy-saving mechanism was involved in the evolutionary transition from non-organic to organic. Although origin-of-life conditions are beyond the scope of this paper, it is suggested that an energy-saving attractor was foundational to the early-evolution of proto-cell coupling, aggregation, differentiation, and sorting.

*1.1    The cell-sorting model*

The model presented involves cell aggregates exhibiting a narrow range of metabolic capacities. Cells exploit an energy-saving mechanism and sort themselves during migration into sub-groups. The model used here to describe this sorting is based on a model of peloton dynamics (Trenchard et al., 2015), and simulates primordial forms of cellular attraction, aggregation, sorting, and patterns of collective motion.

Generally, a peloton is a group of cyclists. More precisely, a peloton is defined as a group of cyclists coupled by the energy-saving benefits of drafting (Trenchard et al., 2014). Drafting occurs when cyclists ride behind and at side angles to others in a region of reduced air-pressure; this saves energy for riders who follow (e.g. Belloli et al., 2016). Thus, in a peloton, where drafting allows for significant energy savings, weaker cyclists can hold speeds set by stronger cyclists that would be unsustainable when travelling in isolation.

Pelotons oscillate between different collective behavioral phases, quantitatively identified by equivalent thresholds of collective speed, power, or metabolic output; and qualitatively identified by easily observed changes in collective geometries (Trenchard et al., 2015). These behavioral dynamics can be modelled by three basic physical parameters (Trenchard et al., 2015), which can be applied analogously across orders of magnitude, including microscopic cell aggregations.

The three basic physical parameters are:

- a narrow range of cells' maximal sustainable output capacities (speeds or metabolic function) among cells within an aggregation;
- an energy-saving mechanism and energy-saving quantity that permits following cells to expend less energy while maintaining the pace of leading cells;
- a variable pace or rate of expended energy (described equivalently as a "starvation-level" which increases as cells lose energy), controlled by the experimenter to simulate the pace set by leading cells.

These parameters are incorporated into a computer algorithm and simulation. By varying the cell starvation-level, cells self-organize according to whether they are above or below certain output thresholds. These thresholds define specific phases, to be discussed more subsequently. Thus, by varying the cell starvation-level, aggregates may migrate, divide and combine, elongate or widen, and deposit weakened or expired cells in the path behind.

*1.2    Energy-saving mechanism*

It is proposed that an energy-saving mechanism was involved in cell or proto-cell attraction in early evolutionary environments. This is conceivable because of both the ubiquitous presence of such energy-saving mechanisms in nature (Trenchard and Perc, 2016), and the existence of such mechanisms in non-organic sedimentary systems, as discussed (Wang et al., 2014).

Reductions in energy requirements among biological systems are frequently described in terms of percentages. For example, Lissaman and Schlossenberger (1970) calculated energetic savings of up to 70% per bird in a flock of 25 geese in flight. Individual geese flying in a flock of 55 have been shown empirically to reduce power requirements by 36% (Hainsworth, 1987). By exploiting hydrodynamic drafting, grey mullet fish can reduce tail beat frequency by 28.5% per fish relative to conspecifics swimming alone (Marras et al., 2015). Spiny lobsters migrating in queues reduce drag by up to 65% per lobster compared to migrating individuals (Bill and Herrnkind, 1976). Similar energy savings were hypothesized and extrapolated to be present among extinct trilobite species, resulting in similar savings (Trenchard et al., 2017). Ducklings swimming together on water surfaces reduce metabolic costs by up to 64% per duckling as compared to those swimming alone (Fish, 1994).

At the human scale, energetic reductions are similar in percentage savings. Cyclists in pelotons of eight or more riders who are exploiting aerodynamic drafting have been shown to reduce oxygen consumption requirements up to ~39% (McCole et al., 1990). Recently, Blocken et al. (2018) found that cyclists in optimal energy-saving positions in a peloton of 121 cyclists moving at 54 kmh$^{-1}$ reduce drag up to 95% compared with a cyclist travelling the same speed in isolation and in the same conditions (i.e. road gradient, wind etc.), as shown in Fig. 2(b).

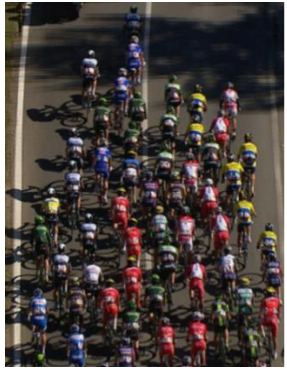      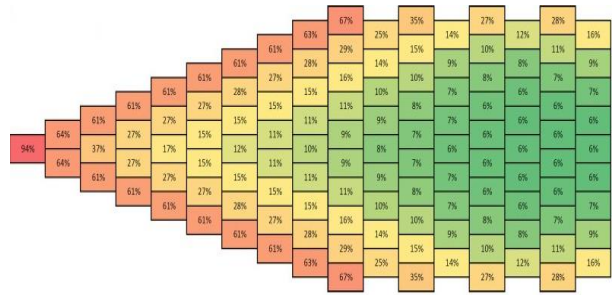

(a)                                      (b)

**Fig. 2** (a). Peloton (reprinted with permission; from Figure 1(b), Trenchard et al., 2017). (b). Reductions in drag forces for a peloton. Each rectangle represents a cyclist, with the group travelling from right to left. Percentages indicate the drag force relative to a cyclist traveling in isolation at 54 kmh$^{-1}$, with greatest drag reductions in centre-rear regions (Blocken et al., 2018, from Fig. 22; Creative Commons licence).

As a natural consequence of reduced energy requirements in a peloton, cyclists fatigue more slowly in a group than they would when isolated (Gaul et al., 2018). This may be presumed a ubiquitous effect among organisms. In turn, we would expect groups enjoying these advantages to either migrate faster than solitary travellers or enjoy periods of migration at much higher speed. This phenomenon is a recognized feature of peloton dynamics (Olds, 1998; Wolfe and Saupe, 2017) and has also been observed among bacteria (Cisneros, 2007), spermatozoa (Moore and Taggart, 1995), and migrating *Dictyostelium discoideum* slugs (e.g. Bonner, 1967, p. 95-96; Inouye and Takeuchi, 1979; Kuzdzal-Fick et al., 2007), as shall be further discussed.

In pelotons, cyclists often take turns sharing anterior positions of highest aerodynamic drag (Olds, 1998; Trenchard et al., 2014; Wolf and Saupe, 2017). Thus, riders take turns in the lead position, setting the group pace. In this position, the lead rider expends greater energy than would be expended anywhere else in the peloton: it is the highest-cost position. Wolf and Saupe (2017) demonstrated a 10% improvement in race time for two riders cooperating in this way compared to each riding in isolation. Olds (1998) reported that when riders alternate between lead and drafting positions, the mean velocity of a peloton rises rapidly as the group size increases from one to five, then tends to flatten with comparatively small increases in velocity up to about 20 riders.

Similar position-changing behavior, explainable in terms of highest-cost-position sharing, has been observed in Bald ibis' in flight (Voelkl et al., 2015), and other birds (Andersson and Wallander, 2004, and citations therein). Similar front-position trading has been observed among spiny lobsters (Bill and Herrnkind, 1976), various fish (Domenici et al., 2002, and citations therein), spiderlings migrating in single file (Reichling, 2000), and has been suggested to occur among extinct trilobites (Trenchard et al., 2017); however, in these cases no clear explanation for the position-changing dynamics has been articulated.

As cyclists in pelotons vary their power output, two main collective geometrical formations emerge as a function of collective output: single-file, or stretched formations, which occur at high collective output; and compact, roughly circular, formations which occur at low and intermediate collective outputs (Trenchard et al., 2014). Compact formations may exhibit varying dynamics, including: 1. high-frequency and high-magnitude positional change, involving frequent position-sharing noted in the foregoing; and 2. low-frequency and low-magnitude positional change (Trenchard et al., 2014). Periods of high-frequency and high-magnitude positional change occur at intermediate speeds. Here, the peloton exhibits convective behavior in which "heating" cyclists advance along peloton perimeters while "cooling" cyclists move backward within the peloton. Periods of low-frequency and low-magnitude positional change (position-locking) in compact formations appears to occur at lower outputs (Trenchard et al., 2014) and during temporary relaxations in pace after high-output efforts (Trenchard, 2010).

When comparing human to non-human systems, reasonable criticisms arise because of confounding factors deriving from human agency, including the capacity to both act on psychological motivations and introduce strategy that deviate from the dictates of energy-saving dynamics. Nonetheless, precise models of peloton dynamics have been produced which discount human strategy and psychological factors to focus exclusively on basic physical and physiological and energetic principles (Trenchard et al., 2014; 2015; Trenchard, 2015).

*1.3  Cell metabolic output ranges*

The proposed "cell peloton" model incorporates a specific range of maximal cellular metabolic capacities, described as the "output-range." In general, where body-lengths are reported in the literature, but maximal capacities are not, maximal output capacities may be estimated by assuming a correspondence between length and speed (Meyer-Vernet 2015). Thus, the output-range approximately corresponds to the range of cell body-size (more specifically, body-length) exhibited in a given cell aggregation.

Cellular output across a specific narrow range is a component of the "variation range hypothesis" (Trenchard, 2015; developed more fully in Trenchard and Perc, 2016, and Trenchard et al., 2017). This hypothesis posits that when a collective set of conspecific organisms exploit an energy-saving mechanism, their body-sizes fall within a range, as a percentage, that roughly corresponds with the energy-saving quantity, as a percentage. This is thought to occur because weaker conspecifics can maintain the pace of stronger ones by exploiting the energy-saving mechanism, particularly during migration; whereas those individuals outside (weaker than) the corresponding equivalent energy-saving quantity are left behind to become separated permanently from the group, either to form isolated sub-groups, or as solitary individuals which will not reproduce.

The hypothesized body-size range and its equivalent output-range may be modulated by the length of the collective. Because of continuous collective movement and positional adjustments, it is difficult for individuals to hold optimal positions. Individuals may drift into zones of reduced energy-saving or temporarily drift outside energy-saving zones altogether. In these circumstances, individuals may exceed their capacity to hold the pace, and drift backwards within the collective. Hence, if the aggregate decelerates to a sustainable speed before weaker, or temporarily weakened, individuals separate from the group, they remain integrated. Since pelotons are observed to exhibit natural oscillations in collective speed, our model dictates that such decelerations are inevitable (Trenchard et al., 2014; 2015). The larger and longer the collective, the more time there is, and the greater the number of opportunities there are, for weaker individuals to remain in contact with the group, as shown in Fig 3. Thus, in a larger and longer collective there is potentially a greater range between the strongest and weakest cells. No known empirical studies exist to quantify this effect.

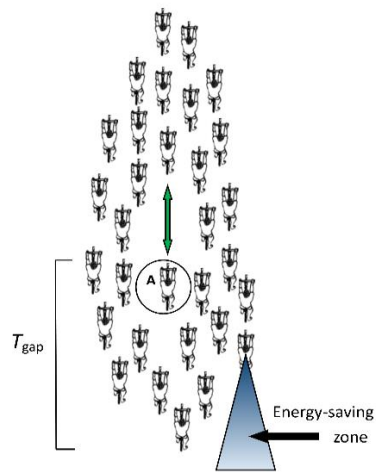

**Fig. 3**. Illustrating a peloton separation. Behind each rider is a zone of energy-saving that diminishes in magnitude with angle, and distance up to about 3 m, indicated by the fading blue triangle. The darker blue region, closest to the cyclist, indicates greater drafting magnitude, and light blue represents reduced drafting magnitude. Here, Rider A's maximum speed is not enough to hold the pace of the leader even by drafting, and A decelerates relative to the group as faster riders go past. The green vertical arrow shows increasing separation between rider A and the rider ahead. $T$gap is the time required to travel backward the distance between rider A and the rear-most optimal drafting position in the peloton. $T$gap increases as the number of riders increases behind rider A (i.e. if A is closer to the front, or if the peloton consists of more riders) or if the peloton stretches. Thus, A remains part of the peloton if the pace slows before $T$gap expires (modified from Trenchard et al., 2014, Fig. 8, reproduced with permission).

*1.4 The relationship between cell output capacity, pace, and energy-saving quantity*

Under the present model, cell outputs are determined by two factors: 1. individual cell maximum output capacity, an intrinsic property of each cell (this can change with fatigue, but this is not modelled here); 2. the ratios of each cell's *current* outputs to their *maximum* output capacity at any given moment. The current output of a following cell, as attenuated by the energy-saving quantity, is determined by the pace (speed) of the leading cell, or by a cell's own pace if it is the leading cell (the "pacesetter"). Here, the pacesetter speed is controlled externally by the experimenter.

These factors are incorporated into a fundamental coupling equation (Trenchard et al., 2014):

$$CCR = \frac{P_{\text{front}} - (P_{\text{front}} * (1-d))}{MSO_{\text{follow}}}, \qquad (1)$$

where *CCR* is the "cell convergence ratio" which describes the relationship between two cells coupled by an energy-saving mechanism, and is readily generalizable to aggregates of unlimited size; $P_{\text{front}}$ is the power output of the leading cell, which may be given by equivalent individual speed in a constant environment (i.e. environments in which speed correlates at all times with power output), and is equivalent to other output metrics such as cell oscillations or temperature fluctuations that can be shown to be driven by energy consumption; *d* is the coefficient of drafting, which is the ratio of the output of agent *x* in an energy-saving position to the output of agent *y* in a high-cost (non-drafting) position; 1- *d* is the energy-saving quantity, and may be expressed as a percent: (1 – *d*) / 100; $MSO_{\text{follow}}$ is the maximum sustainable power output, speed, or metabolic rate of the cell following in the energy-saving wake.

The dividend in equation (1) may be written $P_{\text{front}} * d$. If $d = 1$, meaning there is no energy-saving, *CCR* represents an uncoupled circumstance and is simply the ratio of the output of the leading cell to the maximum of the following cell. If $d = 0$, representing 100% energy-saving, then *CCR* = 0, an impossible circumstance in natural systems which necessarily involve friction or gravity or both.

*1.5 Cell output phases*

From equation (1) we infer three distinct phase-states according to cells' relative energetic vigor, as follows:

> a. **dying or dead phase**: when the leader's pace is too high for following cells to sustain, even by exploiting the energy-saving mechanism, the following cells are forced to decelerate and move backward relative to others; and, unless the pace slows before the following cells slip to the end of the group, they will be separated from the group; these are depicted as **yellow** cells in the figures;
>
> b. **suffering phase**: when the leader's pace is too high for following cells, but is offset sufficiently by the energy-saving mechanism, weaker cells stay within the aggregate but do not move into high-cost leading positions; these cells are depicted as **red** cells in figures;

      c.    **active phase**: when following cells' output capacities exceed the output required to match the pace set by cells in leading positions, the followers will always be strong enough to pass those ahead and assume leading positions in which they can then set the pace; these cells are depicted as **green** in the figures.

These phases can be identified by the following inequalities:

$$\text{In all cases:} \qquad CCR > 1 \text{ (dying or dead; yellow)} \qquad (2)$$

$$\text{If } MSO_{\text{follow}} < P_{\text{front}}: \qquad d < CCR < 1 \text{ (suffering; red)} \qquad (3)$$

$$MSO_{\text{follow}} > P_{\text{front}} \text{ (active; green)} \qquad (4)$$

Inequality (3) applies when following cells' maximum capacity is less than the pace set by the leading cells. In this case, the energy-saving quantity effectively increases the followers' maximal output, allowing them to sustain the pace of the leaders. In other words, if these following cells were not in an energy-saving zone, they could not sustain the pace of the faster leader cell and would therefore enter a dying/dead phase. If $MSO_{\text{follow}} > P_{\text{front}}$, following cells do not require the energy-saving boost of drafting, since their output capacity matches or exceeds that of the leader; hence, they are always "active", as indicated by inequality (4).

## 2. Methods

*2.1 Simulating collective cell behavior*

For micro-organisms moving in viscous environments, forces such as diffusion, cohesion, surface tension, and viscosity have large effect, compared to the great effect that gravity and inertia have on more massive organisms (Bonner, 1988, p.107). These factors represent complex fluid dynamics, and the mathematical solutions of motion for micro-organisms are highly complex (e.g. Subramanian and Nott, 2012) and a "massive undertaking" (Ishikawa et al. 2006, p. 120). Moreover, where in our simplified model we consider decelerating cells relative to each other, we may anticipate further mathematical and simulation complexities in a more realistic model.

For *D. discoideum,* cell–cell and cell–substrate forces include adhesion, resistance to deformation, the locomotive force that cells apply to neighbors or to the substrate, and the drag of extracellular fluid (Palsson and Othmer, 2000). Arguably, if the present model purports to account for *D. discoideum* energy-saving and consumption, then our simulation ought to account for these factors with some analysis of the various specific mechanical forces involved.

However, for the purposes of our theoretical framework, it is unnecessary to consider the specific complex fluid dynamical and related factors of *D. discoideum* cell populations. Fundamentally, our objective is to model the output phases of general peloton behavior as described by equations (1) to (4). These dynamics can then be generalized analogously to behaviors observable in cell populations and to *D. discoideum slugs*. To this end, our model relies on broadly scalable, dimensionless proportions and percentages, given in terms of a range of metabolic outputs, and varying energy-savings percentages. Similarly, variable speeds or power-outputs represent dimensionless proportions of an exhaustion or starvation threshold.

Thus, for our simulations, we have retained most of the parameters and simulation code relied upon by Trenchard et al. (2015) (see Appendix), including the power-output and deceleration equations. Aside from modifications to the code to identify cell-output phases, the only parameter necessary for our analysis that is specific to *D. discoideum* is the cell-size range, which is used to derive a range of cell metabolic outputs and to infer an energy-saving quantity.

*2.2 Determining cell metabolic capacities, and an energy-saving quantity*

Although *D. discoideum* cells can achieve speeds of 40 µmin$^{-1}$ (Potel and MacKay, 1979), we found no literature reporting maximum speeds for a set of individual cells when migrating in isolation, which would allow us to establish a range of maximum outputs for an aggregate. However, this range can be approximated. As noted, the range of cell metabolic outputs (*MSO*) and the energy-saving quantity are hypothesized to be related. Thus, knowing one allows us to approximate the other (Trenchard and Perc, 2017). Additionally, since body-size correlates to motile speeds, by knowing the former we can determine the latter (Mayer-Vernet 2015). And, since speed is one measure of cell metabolic output, if we know maximal *D. discoideum* cell speeds, we can determine range of cell metabolic capacities.

Bonner and Frascella (1953) reported cell-size diameters between ~5.2 µ to ~12.1 µ (~57%, where range% = (max-min) / max), for migrating cells, and ~4.7 µ to ~11.6 µ (59%) for aggregating cells. These figures suggest a variation of 58% as a plausible hypothetical *MSO* range; and they also imply an acceptable range in cell length of between 50 and 120 (no units, since differences are scaled proportionately).

Bonner et al. (1971) reported cell-size length ranges as great as 67% (their Fig. 5). Thus, taking into consideration Bonner and Frascella's (1953) variation of 58%, and applying an approximate equivalence between the energy-saving-quantity and cell-size, we may hypothesize energy-saving between 58 and 67% (mean = 62%) for our simulations.

*2.3 Initializing the simulation*

Agent-based computer modelling platform Netlogo 5.1 (Wilensky 1998; 1999) was used to simulate collective cell dynamics, generating a wide range of possible collective, two-dimensional cell aggregate shapes and dynamics. See the Appendix for the simulation code and details of the simulation protocol.

We rely upon the peloton model of Trenchard et al. (2014) to demonstrate analogous cell behaviors. In the Trenchard et al. (2014) model, equation (1) and related mathematical modifications were introduced into Ratamero's (2013) peloton cohesion, alignment, and separation algorithm, and Ratamero's (2013) algorithm for power output reductions due to drafting based on Olds (1998) and Kyle (1979). The modeling principles of collision avoidance, velocity matching, and flock centering are traced to Reynolds (1987).

Cells are randomly assigned a maximal sustainable output (*MSO*) of between 50 and 120. To initialize the simulation, cells are plotted randomly on a grid. When running, the simulation computes all cell *x* and *y* coordinates to determine the centroid, or the centre of mass of the network of cells. When the simulation begins, the simulation algorithm directs the cells to move toward the centroid while maintaining a minimum distance from each other. This represents the general tendency for cells to both attract to low-energy (energy-saving) regions, and to aggregate. After aggregating toward the centroid, cells in the simulation move generally horizontally along the x-axis with small, average random accelerations; there is also random movement along the y-axis, but it is proportionately smaller.

In the simulation, output in terms of speed, or starvation-level, is a control variable which is set for all cells. Accordingly, every cell responds differently depending on whether they are in energy-saving positions or not, and depending on own maximum sustainable speeds. Here, since speeds are assumed to apply within constant conditions (i.e. soil, humidity, temperature, wind, ground slope, etc.), speed or output-level is conceptually equivalent to starvation, or the process of cellular consumption of energy; therefore, we use "output-level" and "starvation-level" interchangeably.

The simulation algorithm was set to approximate initial slug shape. When running, however changes in slug shape self-organized according to collective cell speeds.

*2.4 Simulation weaknesses*

This model is not useful for confirming or predicting specific cell motion or speeds: it merely demonstrates principles of analogous peloton behavior that we generalize to the behavior of *D. discoideum* and to cell populations. Additionally, other simulation parameters are simplified, including the space between simulated cells, and those that determine 2-dimensional slug shape. These oversimplifications result in distortions in the simulation, including over-elongation of slugs, and low cell density compared to actual *D. discoideum* in which minimum densities are required for cell aggregation and cell-type differentiation within slugs (Town et al., 1976).

Nonetheless, future work should introduce cell-specific forces and measurements to accurately test the principles of this model. A starting point for more precise models that may test the principles set out here and which account for precise drag forces and related factors, are those models by Vasiev and Weijer (2003), Palsson and Othmer (2000) and Dallon and Othmer (2004).

2.5. *Experimental design*

To demonstrate the effects of an energy-saving mechanism on cell sorting within aggregations, simulation tests were conducted applying six energy-saving regimes: 0%, 25%, 50%, 62%, 75% and 95%. 0% represents no drafting. 95% represents the generous drafting demonstrated in Fig. 2(b) (Blocken et al., 2018), where cyclists are in a very high peloton speed of 54 kmh$^{-1}$. 62% is the energy-saving quantity we determined as a reasonable approximation from the reported cell size ranges of *D. discoideum*. 100% energy-saving is excluded because it is an impossible occurrence in natural systems involving viscosity, friction and gravity.

Experiments were initiated with output-level set such that all cells were active (green) at relatively low output (here output-level 4). Every 500 time-steps, the output-level was increased by 1 to simulate gradual starvation. By varying the output-level, the output phase occupied by each cell is determined; i.e. whether cells were in a dying or dead, suffering, or active phase. Changes in cell color proportions were recorded, and individual tests were stopped when yellow cells approached 100% of the population, regardless of corresponding starvation-level, as shown in Fig. 4.

A slug size of 1000 cells was deemed optimal for the tests involving variable energy-saving percentages, because it accommodated our limited computational capacity while effectively simulating the collective behavior under study. Further tests involving slugs of 5000 cells were conducted to demonstrate the effects of directing weaker cells to migrate faster than stronger cells. This comparatively large number of cells permitted the formation of several slugs of different size, as shown in Fig. 7.

## 3. Results

*3.1 Simulation results and cell sorting*

As a general observation, when simulated cells underwent increasing output-levels, eventually they reached the dying threshold. At that point cells were forced to decelerate until their output fell below the dying threshold. If dying cells were in non-drafting positions when they encountered the dying threshold, they could resume below-threshold outputs by moving backward into an energy-saving region, at which point they could re-establish forward movement. However, if the output-level was sufficiently high, dying cells did not recover by moving into energy-saving regions, and continued to decelerate and shift their positions backwards to the posterior of the slug, resulting in increasing slug length, or sloughing of cells from the rear. In this way, our simulations effectively demonstrate the peloton behavior illustrated in Fig. 3.

Further, the simulation experiments demonstrated that variation in slug output-levels produce slug composition ratios expected from the peloton model of dynamics. When simulated slugs commenced migration at low outputs, compositions consisted of high ratios of active-to-suffering cells. As output-levels were increased, higher suffering-to-active ratios were observed.

As outputs were increased further, compositions shifted to high suffering-to-dying ratios, as shown in Fig. 4. Fig. 4 also demonstrates that as energy-saving quantities increase, cell life is extended for greater durations and over higher outputs (i.e. cells stay below the dying-dead threshold).

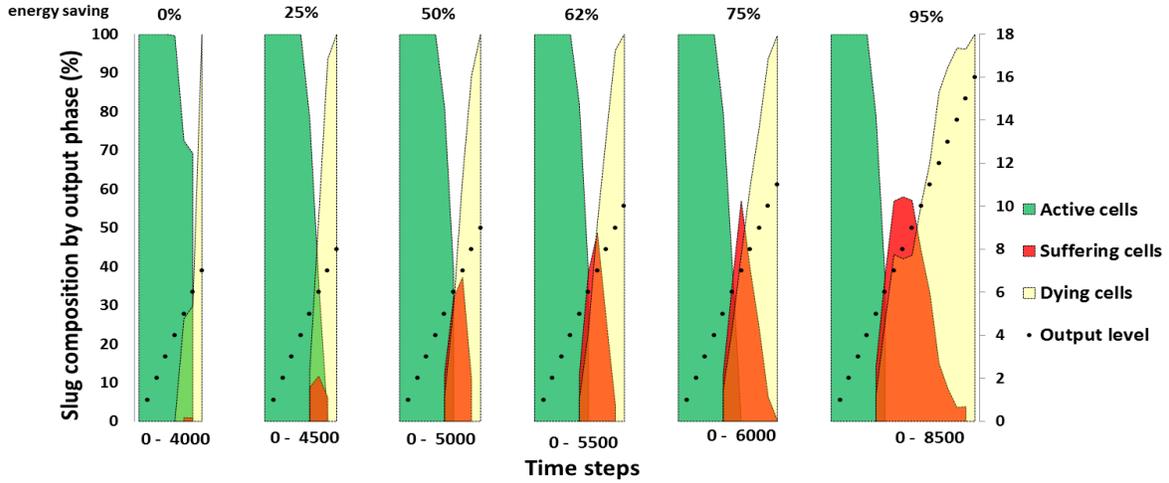

**Fig 4.** Simulated slugs (1000 cells) produce cell-output proportions for three output phases: active (green), suffering (red), and dying or dead (yellow). Slug composition varies as a function of the energy-saving quantity for cells in energy-saving positions and increasing output. Six energy-saving quantities are compared, and outputs were incrementally increased until all cells reached the dying/dead state, the occurrence of which takes longer with greater energy-saving quantities. The colors (online) correspond to phases shown in Figs. 5-8.

A progression of slug compositions for energy-savings of 62%, as a function of increasing outputs, is shown in Fig. 5.

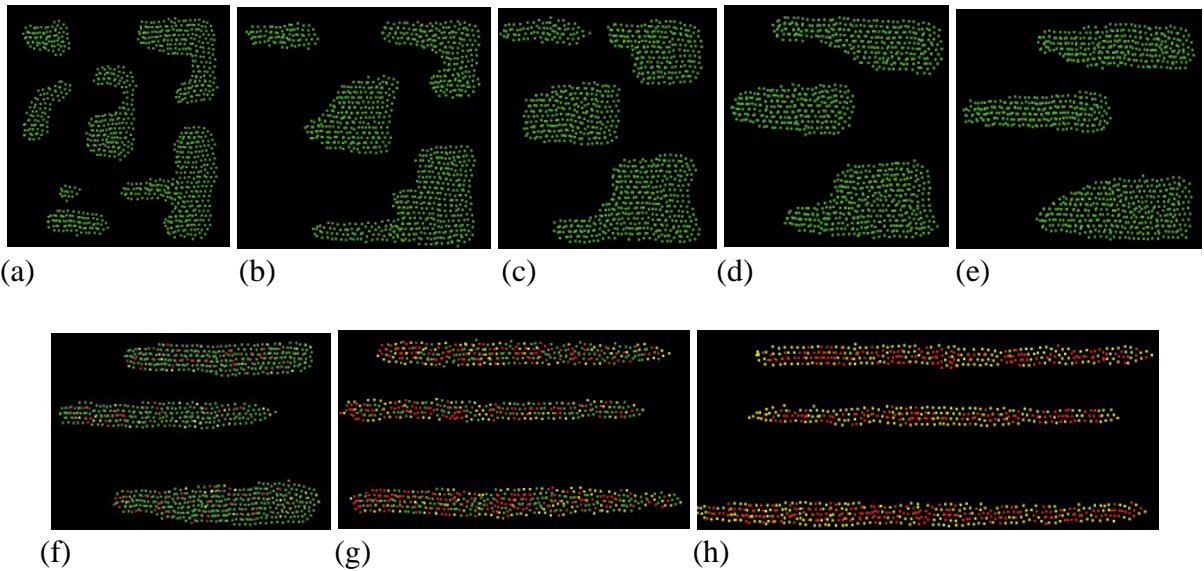

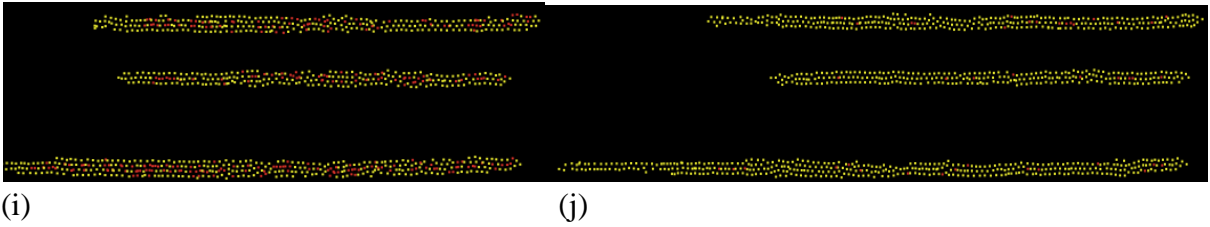

(i)                                               (j)

**Fig. 5.** Slug shape progression and composition changes as outputs are increased incrementally for 1000 cells with energy-saving quantity 62%. Slugs migrate left to right (anterior right).

| Figure | a | b | c | d | e | f | g | h | i | j |
|---|---|---|---|---|---|---|---|---|---|---|
| Output-level | 1 | 2 | 3 | 4 | 5 | 6 | 7 | 8 | 9 | 10 |
| Time steps | 500 | 1000 | 1500 | 2000 | 2500 | 3000 | 3500 | 4000 | 4500 | 5000 |
| Phase proportion (~%active / suffering / dying) | 100 / 0 / 0 | 100 / 0 / 0 | 100 / 0 / 0 | 100 / 0 / 0 | 100 / 0 / 0 | 82 / 12 / 6 | 38 / 39 / 13 | 0 / 59 / 51 | 0 / 26 / 74 | 0 / 0 / 100 |

Also, as predicted by the model, at sufficiently high output-levels, suffering (red) cells and dying/dead (yellow) cells tend to shuffle to the slug posterior, while active cells increasingly occupy the anterior, as shown in Fig. 6.

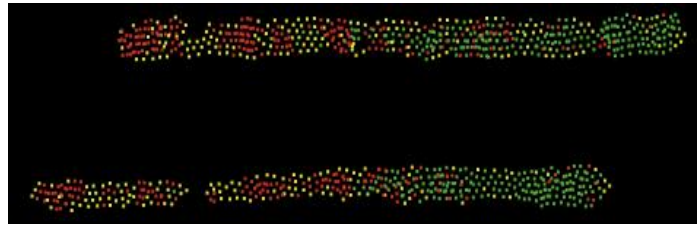

**Fig. 6.** Cell sorting such that active cells occupy slug anterior and suffering and dying cells occupy the posterior. 40.5 % active (green) 33.1% suffering (red) 26.4% dying/dead (yellow). 62% energy-saving at output-level 7, time step 5053: Slug migrates left to right (anterior right). Image was taken moments after separation occurred between lower slug segments at a weak point among decelerating (dying) cells.

Modifying the algorithm slightly by making suffering cells marginally faster than active cells (despite being weaker), leads to the condition in which weaker (red and yellow) cells aggregate in the slug anterior, as shown in Fig. 7. This simulates the known condition in which weaker cells mass in the anterior (Noce and Takeuchi, 1985), while stronger cells recede to the posterior where they conserve energy.

In Fig. 7, the simulation was initialized on a grid of 5000 randomly positioned cells. Typically, these 5000 cells formed several slugs of varying size. Among these slugs, slug composition in terms of cell output phase was consistently observed to be independent of slug-size, a finding consistent with reported observations (Raper, 1940). Repeated simulations revealed that phase composition is conserved across slugs of varying size. The simulations resulted in comparatively elongated slugs, and thus appear perhaps more like *Dictyostelium polycephalum* than *D. Discoideum* (Bonner, 1967, Fig. 12).

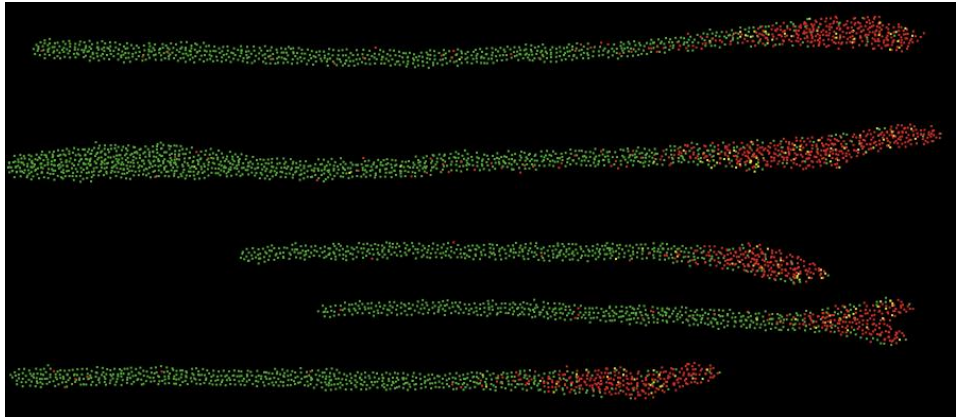

**Fig. 7**. Weaker cells advancing to slug anterior regions. This is achieved by introducing a minor algorithm modification in which suffering (red) cells are given slightly faster speeds than active cells. If suffering cells do not exceed their metabolic threshold, they can achieve faster speeds than stronger, active (green) cells. 5000 cells comprise the five slugs of different size in this image; 62% energy-saving over 2500 time steps, at output-level 6.3. Slugs migrate left to right (anterior right). Composition of each slug, inferred from the recorded total for all 5000 cells: ~69% active (green), ~28% suffering (red), ~3% dying (yellow).

A further minor modification to the algorithm serves to simulate stalk formation. Here, dying cells are directed to the horizontal axial mean of the slug and cease movement, effectively dying and shedding themselves from the posterior of migrating aggregates, as shown in Fig. 8. This simulates the behaviour of slugs of *Dictyostelium mucoroides* or *Dictyostelium purpureum*, which lay stalks as they migrate (Bonner 1967, Fig. 6), or which simply shed cells into the slime stream that pass backward through the slug (Sternfeld 1992); or simulates foreshadowing the fruiting body process when vacuolated cells form stalks (Bonner 1967, p. 67).

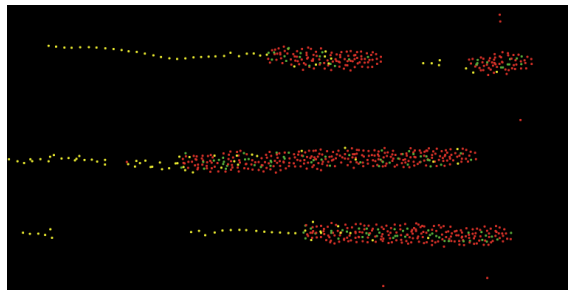

**Fig. 8**. Cell shedding. Simulated vacuolated (dying) cells from the slug body are shuffled to the slugs' rear and shed into trails, simulating stalk formation in select *Dictyostelium* species, or foreshadowing the fruiting body process when vacuolated cells form stalks. Slugs migrate left to right (anterior right).

## 4. Discussion

### 4.1 Model as early evolutionary stage of many organisms

The model demonstrates that reductions in collective energy expenditure permit aggregates to migrate longer and at faster mean speeds than solitary individuals. This expands foraging range and increases access to suitable areas for reproduction, or cell/spore dispersal. Further, the model offers an explanation for certain internal collective cell dynamics, including those of cellular slime mold species other than *Dictyostelium discoideum*. The model also potentially explains aspects of tissue formation within larger organisms, and hints at a precursor to a wide range of botanical species. With respect to slime molds in general, it is conceivable that species evolved at different output phase compositions and in the presence of varying energy-saving quantities and mechanisms, as suggested by Fig. 4.

### 4.2 Applications to D. discoideum

*D. discoideum* begin life as individual amoeba which search independently for food (bacteria) in random directions; when food runs low and cells begin to starve, an aggregation process is triggered (Kessin 2001, p. 1- 4). We suggest that in early-evolutionary conditions, proto-cells were naturally attracted to remain in energy-saving positions, once such positions were encountered during random movement. Being reproductively advantageous, attraction to such positions triggered an evolving tendency for cells to seek such positions, assisted at later evolutionary stages by chemotaxis and other mechanisms.

After aggregating, cells form slugs which "migrate as a bag full of cells" (Bonner, 1994, p.1). Slugs typically contain between about 11,000 and 2.7 million cells (Bonner, 2001). Slugs deposit cells into their slime trails as they migrate (Sternfeld, 1992). Eventually slugs cease migration, initiating cellular processes that perpetuate the *D. discoideum* lifecycle. When the slugs' migration is arrested, a mound of cells develops, under which prestalk cells move downward to form stalks, while prespore cells ascend (Kessin 2001, p. 199) to form a fruiting body from which spores are then shed to begin the lifecycle anew: see Bonner (1967), Kessin (2001) for general descriptions of the *D. discoideum* lifecycle.

Slugs have been observed to increase fitness for several reasons, including: to protect themselves from nematodes (Kessin, 1996) and from noxious environments; and to create a resistant spore, placed for long-distance dispersal (Bonner, 1982; Kessin, 2001, p. 188). Further, Kuzdzal-Fick et al. (2007) demonstrated that when *D. discoideum* slugs deposit cells into the slime stream, the deposited cells consume bacteria in more distant locations than their solitary counterparts. Thus, deposited cells effectively re-enact earlier stages of their lifecycle by aggregating into slugs after starvation begins anew, and then continue to migrate before transforming into a fruiting-body.

In addition to these processes by which *D. discoideum* has adapted to its environment, we suggest that slugs may well adapt in part by showing peloton-like dynamics. We argue that a substantial survival advantage is conferred by the intra-slug cell dynamics that emerge from cell-output heterogeneity in the presence of an energy-saving mechanism. As discussed, these dynamics include collective migration which terminates in the buildup of a cellular mound, which then triggers the production of prestalk and prespore cells. Thus, these dynamics are pre-conditions for the emergence of the fruiting body, which facilitates spore dispersal to new, food-rich environments. We believe the existence of an energy-savings mechanism is validated by various sources of evidence, including: reports that slugs migrate faster than solitary individuals, cell convective dynamics, and adhesive forces.

Further, we argue the present model provides one explanation for the posterior/anterior cell proportions and why these proportions may differ among species of cellular slime molds, such as the approximate 80:20 ratio for *D. discoideum*, but where no similar posterior/anterior cell ratios are observed among *Polysphondylium* or *D. minutum* (Morrissey, 1982, p. 426-427). The present model also explains the robust boundary between posterior and anterior regions, and the observations that the 80:20 proportions in *D. discoideum* are invariant over slug size (i.e. the number of cells within slugs) (MacWilliams and Bonner, 1979).

*4.3   Evidence for an energy-saving mechanism in D. Discoideum*

Evidence for the presence of an energy-saving mechanism is derived from three primary observations:

1. Average slug speeds that are higher than speeds of solitary cells;

2. Cells that circulate or trade-off positions in the anterior region of the slug;

3. Cellular adhesive forces.

*4.4   Slug speeds higher than isolated cells*

Kuzdzal-Fick (2007) reported that solitary amoeba "…move a great deal more slowly and travel much shorter distances than slugs" and referred to single cells moving at 9.8-14.8 µm/min on agar, citing Rifkin and Goldberg (2006). Varnum and Soll (1984) reported average motility rates (speeds) of 50 individual cells to be between ~4 µm/min and ~ 11 µm/min, depending on available cyclic adenosine monophosphate (cAMP) concentrations. Similar mean speeds for individual amoeba were reported by Fisher et al. (1989).

In contrast, slugs travelling on agar move at a speed of 1-2 mm/h (~17 – 34 µm/min) (Raper, 1940). Breen et al. (1987) reported typical slug speeds of 30 µm/min – at least twice that of solitary amoeba. Slug speeds vary proportionately to slug size and length (Bonner et al., 1953; Inouye and Takeuchi, 1979). Savill and Hogeweg (1997) demonstrated by simulation experiments that amoeba aggregations moved as much as three times faster than solitary cells. During aggregation, individual cells cover distances up to 1 cm (Kessin, 2001, p. 2), whereas slug migrations cover distances of 10-20 cm (Kessin, 2001, p. 171) over durations of up to 10 days (Slifkin and Bonner, 1952).

The higher speed of the cell aggregate when compared to solitary cells suggests an energy-saving mechanism that permits cells in optimal positions to reduce their output. The energy-savings is utilized by the aggregate to generate higher average collective outputs for longer durations, consistent with the peloton model and other organisms, as previously discussed.

*4.5 Cell circulation*

Cell circulation is perhaps a subtler validation of the energy-saving mechanism than is higher slug speed. However, it is still compelling when its dynamics are understood. Circulating cells in the slug anterior may permit fatiguing cells to recover in energy-saving zones located behind others in high-cost positions (i.e. non-energy-saving). The process is continual: active and suffering cells below the dying threshold in energy-saving positions advance until they reach non-optimal positions where they exceed the dying threshold, then decelerate and recover while "fresh" individuals take up the pace.

This cell circulatory pattern is described by an integrate-and-fire oscillator whose output increases to a threshold, and then re-sets to a lower output state before recurring through the same pattern (Glass and Mackey, 1979). Certain aspects of collective dynamics can also be described by this oscillator. For instance, at the collective level, this recurrent pattern of outputs is like a convection flow, an observed feature of bicycle pelotons. Here, lines of accelerating ("heating") and decelerating ("cooling") cyclists move in opposite relative directions (Trenchard et al., 2014; Ratamero, 2015). A similar convection pattern has also been modelled for a pair of cooperating cyclists (Wolfe and Saupe, 2017). In pelotons, at low-to-intermediate speeds, convection occurs when faster cyclists advance to the anterior along the peloton periphery where they are unimpeded; whereas fatiguing cyclists fall effectively backwards axially along central regions of the peloton (Trenchard et al., 2014). As noted, this convection effect is determined largely by opportunities for cyclists to recover in energy-saving regions. We infer that the emergence of similar processes in *D. discoideum* implies an energy-saving mechanism. By visual observations of peripheral cell advancement and horizontal axial backward flow, it is apparent from the simulations that the present model involves convection dynamics; however, we have not precisely quantified the phenomenon here; this may be the focus of future studies.

Similar descriptions of *D. discoideum* dynamics include: Bonner (1998, p. 9356), who described a process involving "leader" *D. discoideum* cells at the slug tip that "are constantly replacing one-another; they seem to take turns being at the front end"; Umeda and Inouye (2004, p. 10), who described how "convection flow arises within the cell mass due to the difference in motive force"; Dormann et al. (1996), who described anterior cell rotational patterns as perpendicular to the direction of motion within the raised slug tip; and Siegert and Weijer (1992), who described this motion as helical or a scroll wave. Sternfeld (1992) described cell movement within slugs as a reverse fountain circulation. This process is also apparent in 2-dimensional (one-cell thick) slugs (Bonner, 1998; Nicol et al., 1999), which supports the application of the present two-dimensional simulation.

We also model another process that appears to be related to the convection process: stalk generation during slug migration, which has been observed in slime mold species *D. purpureum*, *D. Mucoroides* and *Polysphondylium* (Bonner, 1967, p. 38, Fig. 6). This can be achieved by making a minor modification to the simulation algorithm, in which dying or dead cells that become isolated (either individually or in small groups) from the back of the slug are impelled to stop moving to simulate their death, thus depositing themselves into the slime-stream, as shown in Fig. 8.

Further simulation experiments without this additional algorithm are required to study the pattern formations of dying cells as they are deposited into the slime stream. One would expect the formations to self-organize without the forcing effect of an additional algorithm, but at present we do not have enough data to draw conclusions regarding the nature and degree of pattern formation among dead cells sloughing into the slime stream under the present simulation.

In addition, further simulation experiments may provide insight into the precursors of apoptosis in embryonic development. Unquantified observations of the simulations suggest that when cell-death is localized individually or in subsets within cell populations, other cells are caused to re-direct their aggregate migration. In this way, directed aggregate migration may be controlled by varying metabolic output, which determines the number and positions of dying cells that block the paths of active ones, forcing active cells to re-route.

*4.6     Adhesive forces*

Adhesive forces are a reasonable candidate for an energy-saving mechanism. This claim is supported by the description of adhesive effects as a means of minimum cell energy expenditure. In terms of *D. discoideum*, cellular adhesion has been referred to as a minimum free-energy condition that permits faster group movement relative to speeds of isolated individuals (Savill and Hogeweg, 1997; Umeda, 1993). Savill and Hogeweg (1997) observed that cell adhesion is critical to slug movement and speed and suggested that cell adhesion has a greater effect on slug speeds than waves of the chemo-attractant cAMP.

Thus, the presence of an existing possible energy-saving mechanism, in the form of adhesive forces, supports the present model in principle, although there may be other, yet unidentified energy-saving mechanisms present.

We note that if our model is conceived to exist in an early-evolutionary stage, and is therefore populated with proto-cells, it is open to the inclusion of energy-saving mechanisms that preceded cell adhesion, since the adhesive mechanism would seem to have evolved to stabilize proto-cell aggregation which was otherwise achieved by randomly attracting toward energy-saving positions.

*4.7    Energy-saving facilitate cell-sorting and anterior/posterior cell differentiation*

Having outlined some evidence for an energy-saving mechanism in *D. discoideum* slugs, we consider how this may have facilitated cell-sorting and differentiation in slugs at an early evolutionary stage, serving as a precursor to fruiting-body formation and spore dispersal.

Anterior cells are referred to as prestalk cells, whereas posterior cells are called prespore cells; this nomenclature arises from the process of fruiting-body formation in the *D. discoideum* lifecycle, wherein anterior cells die and form the stalk of the fruiting-body, while posterior cells ascend to form the head of the spore-laden fruiting-body (Kessin, 2001, p. 189-192).

The boundary between anterior cells and posterior regions is visually discernable and resists disturbances such as tip excision (Raper, 1940) or cutting of slugs and removal of cells by syringe (Meinhardt, 1983). The two regions exhibit quite different dynamics. Bonner (1952, p. 86) observed "that the individual cells each have their own velocity…Only the fastest will be stalk cells and the slowest will be spore cells." Hence, anterior prestalk cells comprise the "engine" of the slug, and prespore cells the "cargo" (Williams et al., 1986). Since they are energetically "high-spenders," move faster than posterior cells (Bonner, 1998; Inouye and Takeuchi ,1979) and possess higher motive forces (Inouye and Takeuchi, 1980), the convection dynamic appears to occur only among anterior cells.

Unlike anterior cells, posterior cells migrate in relatively fixed positions, are passive and tend to conserve energetic resources (Bonner et al. 1955). Compared to anterior cells, posterior cells are also positioned less densely (Takeuchi, 1969; Bonner et al., 1959), and are less adhesive (Yabuno, 1971).

Counterintuitively, despite being energetically more active, weaker cells, in terms of their energetic supply, tend to populate the anterior, whereas stronger cells congregate in the posterior (Noce and Takeuchi, 1985). This is despite anterior cells generally being larger than posterior cells during migration (Bonner and Frascella, 1953), which would suggest greater strength. Arguably, this is inconsistent with a peloton model in which stronger cyclists drive the pace, while weaker cyclists keep pace by drafting behind them, exploiting the energy-saving mechanism.

Despite its counterintuitive nature, the present model well-accommodates such observations that weaker cells are faster. Certainly, in the present model we noted an increasing tendency for the strongest cells to take leading positions *as the speed increases*. As shown in Fig. 4, during relatively high output, at output-level 7 (at output-level 10 all cells were dead), weaker cells were shuffled backwards as increasing speeds pushed them into the dying/dead phase. Consequently, as speeds increase stronger cells come to dominate the anterior, as shown in Fig. 6.

However, in the present model, pacesetters are not necessarily always the strongest cyclists (or cells). Indeed, actual bicycle racing strategy often dictates that weaker team-mates sacrifice themselves for the stronger cyclists. This allows the stronger riders to remain fresh to contest the final sprint or surges in the closing kilometers (Scelles et al., 2018). We are careful here to distinguish human-based volitional strategy from self-organized processes determined by physical parameters. Nonetheless, we can confirm that there is no expectation in actual pelotons that the strongest group members are pacesetters.

Thus, the apparent contradiction here can be resolved if lower aggregate speeds are involved and cells are not driven to their maximums. At low and intermediate speeds (i.e. at low output-levels), the tendency is for roughly homogeneous mixing among active (green) cells. This is shown in Fig. 5(a) – (e) at output-levels in which all cells are active and cell maximum capacities exceed the output-level.

Cell sorting does, however, occur when outputs increase, and cells shift to the suffering and dying phases. Recall that cells are in the suffering phase when cells' maximum output capacity is lower than the pacesetter speed (output-level). These cells cannot sustain the pacesetter speed (output-level) when isolated. They must exploit the energy-saving mechanism to remain within the collective. By definition, these cells are weaker than the active cells, whose maximum outputs are greater than the pacesetter speeds. We can adjust the parameters of the model to accommodate experimental observation, which confirms the existence of a counterintuitive phenomenon in slug migration: the higher speed of weaker cells. The model permits the programming of weaker cells in the suffering phase with higher speeds than their active counterparts, as shown in Fig. 7. So long as these cells do not exceed their own maximum sustainable outputs, a goal supported by being in an energy-saving position, they can advance faster than inherently stronger cells to the anterior of the simulated slug, albeit "suffering" the whole way there.

Along these lines, Bonner (1959) reported that faster cells sorted into higher concentrations during aggregation by moving around slower cells. Arguing against this, Forman and Garrod (1977) suggested that differential cell speeds during migration was unlikely to be the source of cell sorting because cell sorting occurred prior to migration when cell aggregates exist in spherical masses without any directional cell mass movement. Leach et al. (1973), appears to have taken a more compromising view by suggesting that slug migration was not necessary for cell differentiation since sorting could be observed before slug formation, while acknowledging that some sorting may occur during the migration stage and arguing that cells are predisposed by type. Matsukuma and Durston (1979) argued for a combination of differential adhesion and chemotaxis as the sources for cell differentiation.

Later, Bonner (1994, p.3) re-asserted that speed differential "was part of the evidence that there was a sorting out of cells leading to an anterior prestalk zone, and a posterior prespore zone". Thus, there appears to be little consensus on sorting mechanisms (see also Kay and Thompson, 2009). Indeed, the controversy over the mechanisms for prespore/prestalk cell differentiation has been described as "somewhat heated" (Jang and Gomer, 2011, p. 150), and more recent candidates for cell differentiation include morphogen gradients such as chlorinated alkyl phenones, and cell-cycle dependencies (for review of these, see Jang and Gomer, 2011). In terms of cell-cycle dependency, Azhar et al. (2001) (also Baskar et al., 2003) reported that higher calcium levels at time of starvation tend to produce prestalk cells, whereas lower calcium levels produce prespore cells.

The present model is consistent with Bonner's (1959) earlier description of cells that filter by differentials in cell speed. Even if cells sort while within spherical masses during the aggregation stage prior to slug migration, inherent cell speeds are likely to be critical to this process. As well, differences in cell adhesion suggest that cell speeds are altered according to the degree of friction they experience due to adhesiveness, which differs between prestalk and prespore cells (Maree and Hogeweg, 2001).

Another line of inquiry supports the importance of cell speed and metabolism to cell differentiation. Baskar et al. (2003) determined that, at the time of starvation, relative changes in cell calcium levels influence cell vigor which causes the slug to change shape: higher calcium elongates the slug; and lower calcium shorten the slug. Thus, since cell speeds or metabolic capacities do not seem to be ruled out from these considerations, the use of cell speed as a basic factor in cell differentiation is supportable in the present model.

The present model is also consistent with observations that prestalk and prespore cells can convert between themselves (Sternfeld, 1992). Changes in slug speed or fatigue alter the output phase experienced by cells, as shown in Fig. 4. In this way, cells effectively convert from one kind to another.

In terms of the boundary between prestalk and prespore cells and the observation by Raper (1940) that prespore/prestalk proportions recover after slug tip excision, this may be explained by the present model as follows: once the tip is removed, cells that were previously drafting in the active phase suddenly become exposed to high-cost positions and are pushed into the suffering phase, initiating integrate-and-fire convection dynamics that appear to be integral to tip formation. Hence, the tip re-forms without requiring that cells be predisposed to certain types; without morphogen gradients, altered calcium levels, or chemotaxis, as previously discussed.

The present model further suggests that *D. discoideum* slug compositions are finely tuned to allow weaker cells to drive the pace at a convection rate such that the anterior ~20% of the cells are involved in this process. This tells us *D. discoideum* has evolved to optimize a trade-off between slug mean speed and cell position within the slug, such that strong cells are located posteriorly to conserve energy for the culmination stage, while at the same time active anterior cells prolong slug migration for as long as possible but at a lower mean speed than possible if stronger posterior cells took the lead.

Despite the cost in mean slug speed, it has proven selectively beneficial for weaker cells to occupy the anterior, while the strongest cells conserve their energy for the reward of winning the "finish line sprint," the privilege of becoming the fruiting body and spores. We can imagine the converse arrangement – faster cells to the anterior, and weaker in the posterior – as existing in initial stages of *D. discoideum* evolution; if the strongest cells set the pace, slugs may well have migrated farther and faster, expanding foraging prospects; however, this strategy would have left posterior cells too weak to form a fruiting body when the anterior cells started to die. The optimal combination of slug speed and slug composition – weaker (slower) to the anterior, and stronger (faster), to the posterior – has resulted in greater spore dispersal which, in turn, has produced much greater foraging opportunities.

Moreover, under the present model, and as shown Fig. 7, there is no physical impediment preventing anterior cells from drifting farther into the posterior. However, no impediment is required. Indeed, the cells simply engage the integrate-and-fire mechanism and cell-circulation process, the extent of which produces a self-organized boundary between regions. By drifting back into energy-saving zones until their output-level drops below the dying-phase threshold (i.e. they 'recover'), they can resume forward locomotion toward the front-most high-cost positions. According to the convection dynamic, the cycle continues by cells advancing along the slug-tip periphery, where forward motion is unimpeded; they then drop back along interior trajectories within the slug tip, according to the scroll wave pattern suggested by Siegert and Weijer (1992). This dynamic is also consistent with observations of convective patterns in pelotons (Trenchard et al., 2014).

Hence, the boundary is both well-defined and porous, since changes in cell output can drive simulated cells to flow from posterior to anterior regions. This is consistent with observations that there is no physical barrier between regions (Abe et al., 1994), but contrasts with compartment boundaries of *Drosophila,* which are not crossed (Kessin, 2001, p. 177). The present model therefore permits an explanation for the well-defined, but not physically rigid, boundary between anterior and posterior regions. As noted, we suggest that the system engages a finely-tuned trade-off between slug speed – driven by weaker but highly energetic anterior cells – and the conservation of posterior cells' energetic resources. It is sensitive to output changes that might alter the flow of anterior and posterior cells throughout the slug, but it is otherwise robust in maintaining its current conditions.

Ultimately, at early ancestral stages, we suggest it was simply a matter of chance that groups of highly energetic but weaker proto-cells coexisted with stronger, but less energetic proto-cells in heterogeneous groups. So long as stronger cells were not in locomotion at maximal speeds, these weaker but more active proto-cells could naturally push their way to the aggregate anterior. Repeat incidence of this chance formation would trigger selection for its dynamics because they permit stronger, less active, cells to conserve energy for the production of a fruiting body and spore dispersal. Due to the simplicity of the model and its capacity to explain several reported observations of the slime-mold lifecycle without requiring other causative factors of cell differentiation, it is reasonable to suggest it represents a more primordial stage of evolution.

The present model also accommodates the fact that different varieties of cell-sizes, energy-saving quantities, and collective output-levels, coupled with different environmental conditions can lead to varying slug compositions. Such differences are apparent among different slime mold species (Bonner, 1967, Fig. 10).

*4.8    Anterior/posterior proportion invariance over different slug sizes*

As previously noted, in *D. discoideum* slugs there are approximately 80% posterior cells and 20% anterior cells (Raper, 1940; Meinhardt, 1983; Nanjundiah and Saran, 1992; Sternfeld and David, 1981; 1982). About 5 to 10% of posterior cells are "anterior-like cells", located just behind the prestalk/prespore boundary, which share properties of anterior prestalk cells and flow forward through the prespore region into the anterior region during slug migration (Abe et al., 1994; Sternfeld and David, 1982; Dormann et al., 1996). Meinhardt (1983) described intermingled prestalk and prespore cells as producing a "salt-and-pepper" pattern, rather as our simulations produce, as shown in Figs. 5 – 8.

These anterior/posterior cell proportions are independent of slug size (see MacWilliams and Bonner, 1979, for a review). The present model is consistent with this observation in which cells' output phases in terms of identifiable proportions within slugs, are conserved across slug size, for any given output and energy-saving quantity. As Fig. 4 shows, if we vary output and/or energy-saving quantity, then slug composition varies accordingly; but for any given combination of parameters, simulated slug composition remains constant if slug size is varied. This is shown in Fig. 7, in which cells were randomly distributed on an initial grid of 5000 cells for a given set of parameters; these cells aggregated to form five slugs of varying size, yet cell output phase distributions are roughly equal for each simulated slug present (cell proportions were not quantified, but visual inspection confirms approximate phase proportions).

This is easily explained by the present model: because the range of cell-output capacities is narrow (i.e. randomly distributed between 50 and 120), slug outputs are determined by the given pace expressed as proportions of cells' maximum capacities and attenuated by the energy-saving quantity. Cells' relative outputs are a function of this range for any number of cells. Thus slug phase proportions will be constant for any slug size as long as there is a random distribution of cell outputs across the range (we have not tested any other probability distributions, such as a Gaussian). As discussed, we hypothesize that this output range was determined selectively by the capacity of the weakest cells to sustain the pacesetters' speed by exploiting the energy-saving mechanism; i.e. cells too weak to keep up with the aid of the energy-saving mechanism ultimately do not reproduce.

As discussed in the introduction, the simulated narrow range of metabolic outputs is reasonably expected to occur among actual slug cells. The model is thus highly consistent with actual reported observations of invariant slug composition across slug size.

*4.9    Directions for future work*

*D. discoideum* has been the subject of a vast body of research. It is not possible under this preliminary theoretical overview to explore the implications of the model presented on many elements of *D. discoideum* behavior.

However, the model is particularly amenable to confirming the observation that larger or longer slugs move faster than smaller ones, a result not shown in this paper. The present model permits this because larger slugs contain more cells which migrate at higher proportions of their maximums. In turn, this predicts that, on average, larger slugs will move faster than smaller ones. This observation is not demonstrated in the present simulation largely because pacesetter speeds were arbitrarily controlled with the aim of demonstrating the effects of these speeds on cells' energetic phases. A future study should test slug speeds as a responding variable.

Also ripe for exploration under the present model is the effect of temperature on slug composition. When amoeba form slugs they tend to move toward the soil surface in the presence of temperature gradients (Raper, 1940; Bonner et al., 1950; Whitaker and Poff, 1980). Bonner and Slifkin (1949) reported that an increase in temperature leads to a higher proportion of stalk cells. Similarly, Farnsworth (1973) reported an increase in spore cell proportion with decreasing temperature.

There are, however, several mechanical considerations involved in changing temperatures within biological systems generally that may apply to *D. discoideum*; any application of the various considerations to the present model deserve a detailed analysis that is beyond the scope of this paper.

In addition, future work may consider the model in the context of kin-selection and cooperation-cheater theory. Kin selection models involve cooperators which tend to be clonal (genetically related) and cheaters that tend to be chimera cells (genetically unrelated) (Foster et al., 2002; Gilbert et al., 2007; Khare et al., 2009). By contrast, individual-selection models for the apparent altruism exhibited by cellular slime molds depend on intrinsic cell-to-cell trait differences (Zahavi et al., 2018). Under the present model, as part of each cell's strategy to maximize its own fitness, cell cooperation occurs among more energetic anterior pacesetters who occupy high-cost, non-drafting positions. Stronger cells appear to free-ride, or 'cheat', by exploiting the energy-saving mechanism, effectively hitching a ride on the backs of weaker, but more energetic, pacesetter cells. The model presented here is consistent with an individual-selection model, because our model is grounded in intrinsic differences in cells' metabolic capacities. However, a detailed analysis that reconciles the present model with the noted cooperation theories is beyond the scope of this paper.

## 5. Conclusion

We present a novel theoretical model of an early-evolutionary stage of proto-cellular aggregation that may precede the evolution of chemotaxis, genetic and molecular properties in slime-molds. The model has potential application for a variety of slime mold species, botanical processes, and tissue development.

The model is simple and involves three basic mechanical parameters: a range of proto-cellular output capacities, an energy-saving mechanism, and a motile pacesetter speed. Cells are driven to occupy three different energetic phases depending on these three factors (where pacesetter speeds were varied by the experimenter). For real slugs, natural selection determined the optimal proportion of cells that comprise respective energetic phases.

Cell-output phases are identified according to the following criteria: active cells are those whose maximum capacities are greater than the pacesetter speed; suffering cells are those whose maximum capacities are less than the pacesetter speed, but which can sustain the pace of the pacesetter by exploiting the energy-saving quantity; dying or dead cells are those whose maximum capacities are less than the pacesetter speed, regardless of whether they are in energy-saving zones. Simulation experiments were conducted for different energy-saving quantities and output-levels.

The model offers an explanation, in the absence of chemotaxis, for why cells aggregate in the first place: in a primordial environment, proto-cells attracted by random motion toward existing low-energy positions, after which dynamical processes involving a range of proto-cell energetic capacities, in the presence of an energy-saving mechanism, were selected for. Later, chemotactic, adhesive, genetic and other processes evolved to firmly establish optimal slug cell compositions.

The model well-demonstrates observations of anterior/posterior differentiation and the robust division between these regions; observations that slugs can migrate for longer and at higher speeds than individuals can, as vehicles for eventual spore dispersal; and observations that slug cell composition is invariant over slug size. The model well-demonstrates the deposit of cells into the slug slime-stream and the formation of stalks.

Since there is extensive literature on slime mold, and particularly *D. discoideum*, there are numerous dynamics and matters of controversy which we have not attempted to address. This work may therefore be viewed as a little more than a theoretical starting point.

Further work is required to establish typical ranges of the *maximal* capacities of individual *D. discoideum* cells, and those of other species. Further work is required to establish the nature of an energy-saving mechanism and its quantity or quantities. Establishing these basic parameters are first steps in testing the validity of the model presented. The model is amenable to analysis of aspects of slug behavior including that longer slugs travel faster than slower ones, and the effects of temperature on slug cell proportions. Further work may be done to reconcile kin selection and cooperation-cheater theory with the present model.

**Acknowledgments:** The author acknowledges the comments of an anonymous referee and the editing assistance of Demian Seale, whose input helped to improve this manuscript substantially.

**Conflict of Interest**: The author declares no conflict of interest.

# 6. References

Abe T., Early A.., Siegert F., Weijer C., Williams J., 1994. Patterns of cell movement within the Dictyostelium slug revealed by cell type-specific, surface labelling of living cells. Cell 77, 667-699.

Andersson M., Wallander J., 2004. Kin selection and reciprocity in flight formation? Behav. Ecol. 15 (1), 158–162.

Azhar M.O., Kennady P.K., Pande G.O., Espiritu M.I., Holloman W.E., Brazill D.E., Gomer R.H., Nanjundiah V.I., 2002. Cell cycle phase, cellular Ca2+ and development in Dictyostelium discoideum. Intl. J. Dev. Biol. 45 (2), 405-14.

Baskar R.A., Chhabra P.R., Mascarenhas P.R., Nanjundiah V.I., 2003. A cell type-specific effect of calcium on pattern formation and differentiation in dictyostelium discoideum. Intl. J. Dev. Biol. 44 (5), 491-8.

Belloli M., Giappino S., Robustelli F., Somaschini, C., 2016. Drafting effect in cycling: Investigation by wind tunnel tests. Procedia engineering, 147, 38-43.

Blocken B., van Druenen T., Toparlar Y., Malizia F., Mannion P., Andrianne T., Marchal T., Maas G.J. and Diepens, J., 2018. Aerodynamic drag in cycling pelotons: new insights by CFD simulation and wind tunnel testing. J. Wind Eng. Ind. Aerodynamics, 179, 319-337.

Bill R., Herrnkind W., 1976. Drag reduction by formation movement in spiny lobsters. Sci. 193 (4258), 1146–1148.

Bonner J.T., Slifkin M.K., 1949. A study of the control of differentiation: the proportions of stalk and spore cells in the slime mold Dictyostelium discoideum. Am. J. Bot., 36 (10), 727-734.

Bonner J.T., 1952. The pattern of differentiation in amoeboid slime molds. Am. Nat. 86 (827), 79-89.

Bonner J.T., Frascella E.B., 1953. Variations in cell size during the development of the slime mold Dictyostelium discoideum. Bio. Bull. 104.(3), 297-300

Bonner J.T., Koontz Jr. P.G. and Paton D., 1953. Size in relation to the rate of migration in the slime mold Dictyostelium discoideum. Mycologia, 235-240.

Bonner J.T., Duncan Chiquoine A. and Kolderie M.Q., 1955. A histochemical study of differentiation in the cellular slime molds. J. Exp. Zoo. Part A: Ecol. Gen. Physiol. 130 (1),133-157.

Bonner J.T., 1959. Evidence for the sorting out of cells in the development of the cellular slime molds. Proc. Natl. Ac. Sci. 45 (3), 379-384.

Bonner, J.T., 1967. The Cellular Slime Molds. 2$^{nd}$ ed. Princeton University Press, New Jersey.

Bonner, J.T., Sieja W., Hall E.M., 1971. Further evidence for the sorting out of cells in the differentiation of the cellular slime mold Dictyostelium discoideum. J. Embryol. Exp. Morph. 25 (3), 457-465

Bonner, J.T., 1982. Evolutionary strategies and developmental constraints in the cellular slime moulds. Am. Nat. 119, 530-552

Bonner, J.T., 1988. The evolution of complexity. Princeton University Press, New Jersey.

Bonner J.T., 1994. The migration stage of Dictyostelium: Behavior without muscles or nerves. FEMS. Microbio. Lett. 120, 1-8.

Bonner J.T., 1998. A way of following individual cells in the migrating slugs of Dictyostelium discoideum. Proc. Natl. Ac. Sci. 95 (16), 9355-9359.

Bonner J.T., 2001. A note on the number of cells in a slug of Dictyostelium discoideum (Available at: http://dictybase.org/bonner%20paper.pdf).

Breen E.J., Vardy, P.H., Williams K.L., 1987. Movement of the multicellular slug stage of Dictyostelium discoideum: an analytical approach. Development 101 (2), 313-321.

Cisneros L., Cortez R., Dombrowski C., Goldstein R., Kessler J., 2007. Fluid dynamics of self-propelled microorganisms, from individuals to concentrated populations. Exp. Fluids 43 (5), 737–753.

Dallon J.C. and Othmer H.G., 2004 How cellular movement determines the collective force generated by the Dictyostelium discoideum slug. J. Theor. Biol, 231 (2), 203-222.

Domenici P., Ferrari R.S., Steffensen J.F., Batty R.S., 2002. The effect of progressive hypoxia on school structure and dynamics in Atlantic herring Clupea harengus. Proc. Roy. Soc. Lond. B. 269, 2103-2111.

Dormann D., Siegert F., Weijer C.J., 1996. Analysis of cell movement during the culmination phase of Dictyostelium development. Development 122 (3),761-769.

Farnsworth P.A., 1975. Proportionality in the pattern of differentiation of the cellular slime mould Dictyostelium discoideum and the time of its determination. Development 33(4), 869-877.

Fish F., 1994. Energy conservation by formation swimming: metabolic evidence from ducklings. in: Maddock L (ed.) Mechanics and Physiology of Animal Swimming. Cambridge University Press, Cambridge (UK), pp. 193-206.

Fisher P.R., Merkl R., Gerisch G., 1989. Quantitative analysis of cell motility and chemotaxis in Dictyostelium discoideum by using an image processing system and a novel chemotaxis chamber providing stationary chemical gradients. J. Cell Biol. 108(3), 973-984.

Forman D., Garrod D.R., 1977. Pattern formation in Dictyostelium discoideum: II. Differentiation and pattern formation in non-polar aggregates. Development 40 (1), 229-243.

Foster K.R., Fortunato A., Strassmann J.E., Queller D.C., 2002 The costs and benefits of being a chimera. Proc. Roy. Soc. Lond. B. Biol. Sci. 269 (1507), 2357-62.

Forgacs G., Newman S.A., 2005. Biological physics of the developing embryo. Cambridge University Press, New York.

Fortes A., Joseph D., Lundgren T., 1987. Nonlinear mechanics of fluidization of beds of spherical particles. J. Fluid Mech. 177, 467-483.

Hainsworth F., 1987. Precision and dynamics of positions by Canada geese flying in formation. J. Exp. Biol. 128, 445–462.

Gaul, L.H., Thomson S.J., Griffiths I.M., 2018. Optimizing the breakaway position in cycle races using mathematical modelling Sports Eng. 1-14

Gilbert O.M., Foster K.R., Mehdiabadi N.J., Strassmann J.E., Queller D.C., 2007. High relatedness maintains multicellular cooperation in a social amoeba by controlling cheater mutants. Proc. Natl. Ac Sci. 104 (21), 8913-8917.

Glass L., Mackey M.C., 1979. A simple model for phase locking of biological oscillators. J. Math Biol. 7 (4), 339-352.

Inouye K., Takeuchi I., 1979. Analytical studies on migrating movement of the pseudoplasmodium of Dictyostelium discoideum. Protoplasma 99, 289–304.

Inouye K., Takeuchi I., 1980. Motive force of the migrating pseudoplasmodium of the cellular slime mould Dictyostelium discoideum. J. Cell Sci., 41 (1), 53-64.

Ishikawa T., Simmonds M.P., Pedley T.J., 2006. Hydrodynamic interaction of two swimming model micro-organisms. J. Fluid Mech. 568, 119-160.

Jang W., Gomer R., 2011. Initial cell type choice in Dictyostelium. Eukaryotic Cell, 150-155.

Kay R.R., Thompson C.R., 2009. Forming patterns in development without morphogen gradients: scattered differentiation and sorting out. Cold Spring Harbor perspectives in biology, 1 (6), a001503 https://doi.org/10.1101/cshperspect.a001503

Kessin R., 2001. Dictyostelium: evolution, cell biology, and the development of multi-cellularity. Cambridge University Press, Cambridge (UK).

Kessin R.H., Gundersen G.G., Zaydfudim V., Grimson M., Blanton R.L., 1996. How cellular slime molds evade nematodes. Proc. Natl. Ac. Sci. USA 93, 4857-4861.

Khare A., Santorelli L.A., Strassmann J.E., Queller D.C., Kuspa A., Shaulsky G., 2009. Cheater-resistance is not futile. Nature 461 (7266), 980.

Kyle C.R., 1979. Reduction of wind resistance and power output of racing cyclists and runners travelling in groups. Ergonomics 22 (4), 387–397.

Kuzdzal-Fick J., Foster K., Queller D., Strassman J., 2007. Exploiting new terrain: an advantage to sociality in the slime mold Dictyostelium discoideum. Behav. Ecol. (18)2, 433-437

Leach C.K., Ashworth J.M., Garrod D.R., 1973. Cell sorting out during the differentiation of mixtures of metabolically distinct populations of Dictyostelium discoideum. Development 29(3), 647-661.

Lissaman P., Schlossenberger C., 1970. Formation flight of birds. Sci. 168, 1003–1005.

MacWilliams H.K., Bonner J.T., 1979. The prestalk-prespore pattern in cellular slime molds. Differentiation 14 (1-3), 1-22.

Marée A.F., Hogeweg P., 2001. How amoeboids self-organize into a fruiting body: multicellular coordination in Dictyostelium discoideum. Proc. Natl. Ac. Sci. 98 (7), 3879-3883.

Marras S., Killen S.S., Lindström J., Mckenzie D.J., Steffensen J.F., Domenici P., 2015. Fish swimming in schools save energy regardless of their spatial position. Behav. Ecol. and Sociobiol. 69 (2), 219-226.

Matsukuma S., Durston A.J., 1979. Chemotactic cell sorting in Dictyostelium discoideum. Development 50 (1), 243-251.

Meinhardt H., 1983. A model for the prestalk/prespore patterning in the slug of the slime mold Dictyostelium discoideum. Differentiation 24 (1-3), 191-202.

Meyer-Vernet N., 2015. How fast do living organisms move: Maximum speeds from bacteria to elephants to whales. Am. J. Phys. 83, 718-722.

McCole S. Claney K., Conte J., Anderson R., Hagberg J., 1990. Energy expenditure during bicycling. J. Appl. Physiol. 68, 748–753.

Moore H., Taggart D., 1995. Sperm pairing in the opossum increases the efficiency of sperm movement in a viscous environment. Biol. Reprod. 52 (4), 947–953.

Morrissey, J., 1982. Cell proportioning and pattern formation, in: Loomis W. (ed.), The development of Dictyostelium discoideum. Academic Press, New York, pp. 411-449.

Nanjundiah V., Saran S., 1992. The determination of spatial pattern in Dictyostelium discoideum. J. Biosci. 17 (4), 353-394.

Nicol A., Rappel W-J., Levine H., Loomis W., 1999. Cell-sorting aggregates of Dictyostelium discoideium. J. Cell Sci. 112, 3923-3929.

Noce T., Takeuchi I., 1985. Prestalk/prespore differentiation tendency of Dictyostelium discoideum cells as detected by a stalk-specific monoclonal antibody. Developmental Biol., 109 (1), 57-164.

Olds T., 1998. The mathematics of breaking away and chasing in cycling. Euro. J. Appl. Physiol. Occupational Physiol. 77 (6), 492–497

Palsson E., Othmer H.G., 2000. A model for individual and collective cell movement in Dictyostelium discoideum. Proc. Natl. Ac. Sci. 97 (19), 10448-10453.

Potel M.J., Mackay S.A., 1979. Preaggregative cell motion in Dictyostelium. J. Cell Sci. 36 (1), 281-309.

Raper K.B., 1940. Pseudoplasmodium formation and organization in Dictyostelium discoideum. J. Elisha Mitchell Scientific Soc. 56 (2), 241-282.

Ratamero E.M., 2015. Modelling Peloton Dynamics in Competitive Cycling: A Quantitative Approach. Intl. Congress on Sports Sci. Research and Tech. Support Springer Intl. Switzerland, 42-56

Rosa-Cusachs, P., Sunyer, R., Trepat, X., 2013. Mechanical guidance of cell migration: lessons from chemotaxis.  Current Op. Cell Biol. 25(5), 543-549.

Reichling S.B., 2000. Group dispersal in juvenile Brachypelma vagans (Araneae, Theraphosidae). J. Arach. 28 (2), 248–250.

Reynolds C.W., 1987. Flocks, herds and schools: A distributed behavioral model. ACM SIGGRAPH Computer Graphics 21(4), 25-34.

Rifkin J.L., Goldberg R.R., 2006. Effects of chemoattractant pteridines upon speed of D. discoideum vegetative amoebae. Cell Motility and the Cytoskeleton 63(1), 1-5.

Savill N.J., Hogeweg P., 1997. Modelling morphogenesis: from single cells to crawling slugs. J. Theor. Biol. 184 (3), 229-235.

Scelles N., Mignot J.F., Cabaud B., François A., 2018. Temporary organizational forms and coopetition in cycling: What makes a breakaway successful in the Tour de France? Team Performance Management: An International Journal, 24 (3/4), 122-134.

Siegert F., Weijer C.J., 1992. Three-dimensional scroll waves organize Dictyostelium slugs. Proc. Natl. Ac. Sci. 89 (14), 6433-6437.

Sternfeld J., David C.N., 1981. Oxygen gradients cause pattern orientation in Dictyostelium cell clumps. J. Cell Sci. 50 (1), 9-17.

Sternfeld J., David C.N., 1982. Fate and regulation of anterior-like cells in Dictyostelium slugs. Developmental Biol. 93(1), 111-118.

Sternfeld J., 1992. A study of pstB cells during Dictyostelium migration and culmination reveals a unidirectional cell type conversion process. Roux's Archives of Developmental Biol. 201 (6), 354-363.

Subramanian G., Nott P.R., 2012. The fluid dynamics of swimming microorganisms and cells. J. Ind. Inst. Sci. 91(3), 283-314.

Sunyer R., Conte V., Escribano J., Elosegui-Artola A., Labernadie A., Valon L., Navajas D., García-Aznar., J.M., Muñoz J.J., Roca-Cusachs, P., Trepat X., 2016. Collective cell durotaxis emerges from long-range intercellular force transmission. Sci. 353 (6304),1157-1161.

Takeuchi I., 1969. Establishment of polar organization during slime mold development. Nucleic Acid Metabolism, Cell Differentiation and Cancer Growth, 297.

Town C.D., Gross J.D. and Kay R.R., 1976. Cell differentiation without morphogenesis in Dictyostelium discoideum. Nature, 262 (5570), 717.

Trenchard H., Richardson A., Ratamero E., Perc M., 2014. Collective behavior and the identification of phases in bicycle pelotons. Phys. A. Statistic Mech. Appl. 405, 92–103.

Trenchard H., Ratamero E., Richardson A., Perc M., 2015. A deceleration model for bicycle peloton dynamics and group sorting. Appl. Math. Comput. 251, 24–34.

Trenchard H., 2015. The peloton superorganism and protocooperative behavior. Appl. Math. Comput. 270, 179–192.

Trenchard H., Perc M., 2016. Energy saving mechanisms, collective behavior and the variation range hypothesis in biological systems: A review. BioSystems 147, 40–66

Trenchard H., Brett C., Perc M., 2017. Trilobite 'pelotons': Possible hydrodynamic drag effects between leading and following trilobites in trilobite queues. Palaeontology 60, 557-569.

Trenchard, H., 2010. Hysteresis in competitive bicycle pelotons. Complex Adaptive Systems — Resilience, Robustness, and Evolvability: Papers from the AAAI Fall Symposium FS-10-03

Umeda, T., 1993. A thermodynamical model of cell distributions in the slug of cellular slime mold. Bull. Math Biol. 55, 451-464

Umeda, T. and Inouye, K., 2004. Cell sorting by differential cell motility: a model for pattern formation in Dictyostelium. J. Theor. Biol. 226 (2), 215-224.

Varnum B., Soll D.R., 1984. Effects of cAMP on single cell motility in Dictyostelium. J. Cell Biol. 99 (3), 1151-1155.

Vasiev B., Weijer C.J., 2003. Modelling of Dictyostelium discoideum slug migration. J. Theor. Biol. 223 (3), 347-359.

Voelkl B., Portugal S.J., Unsöld M., Usherwood J.R., Wilson A.M., Fritz J., 2015. Matching times of leading and following suggest cooperation through direct reciprocity during V-formation flight in ibis. Proc. Natl. Ac. Sci. 112 (7), 2115-2120.

Wang L., Guo Z.L., Mi J.C., 2014. Drafting, kissing and tumbling process of two particles with different sizes. Computers & Fluids 96, 20-34.

Weijer C.J., 2009. Collective cell migration in development. J. Cell Sci. 122 (18), 3215-3223.

Whitaker B.D., Poff K.L., 1980. Thermal adaptation of thermosensing and negative thermotaxis in Dictyostelium. Exp. Cell Res. 128 (1), 87-93.

Wilensky U., 1998. Netlogo Flocking model. http://ccl.northwestern.edu/netlogo/models/Flocking. Center for Connected Learning and Computer-Based Modeling, Northwestern University, Evanston, IL.

Wilensky U., 1999. Netlogo. http://ccl.northwestern.edu/netlogo/. Center for Connected Learning and Computer-Based Modeling, Northwestern University, Evanston, IL.

Williams K.L., Vardy P., Segel L.A., 1986. Cell migrations during morphogenesis: some clues from the slug of Dictyostelium discoideum. Bioessays 5(4), 148-152.

Wolf S., Saupe D., 2017. How to Stay Ahead of the Pack: Optimal Road Cycling Strategies for two Cooperating Riders. Intl. J. Computer Sci. Sport 16(2), 88-100.

Yabuno K., 1971. Changes in cellular adhesiveness during the development of the slime mold Dictyostelium discoideum. Development, Growth and Differentiation 13 (3),181-190

Zahavi A., Harris K.D., Nanjundiah V., 2018. An individual-level selection model for the apparent altruism exhibited by cellular slime moulds. J. Biosci. 43(1), 49-58.

3333

# Appendix

**Simulating cell population dynamics using Netlogo 5.1.3. Modified from Trenchard et al. (2015); incorporating cohesion/separation and drafting algorithm from Ratamero (2015).**

___________________________________

*1.0. Properties given to agents, called turtles in Netlogo.*

```
turtles-own
 [speed
  neighborhood
  MSO
  pcr
  draft-coefficient
```

*1.1. Acceleration in x and y-coordinates.*

```
  x-acc
  y-acc
```

*1.2. Power-output reduction when threshold exceeded.*

```
  p_reduction
```

*1.3. Value of power output reduction required to bring cell from max to threshold.*

```
  p1
```

*1.4. Speed after power-to-speed conversion.*

```
  slowing-speed
```

*1.5. Nudge dying cells to horizontal mid-line.*

```
  average-Y ]
```

*2.0. Variables (except for 2.1 to 2.3, all variables were fixed for the experimental protocol in this paper).*

*2.1. Number of cells.*

```
  peloton-size
```

*2.2. Set variable drafting coefficient. This is a primary experimental variable. Six coefficients were considered in this paper.*

```
  draft-coefficient
```

*2.3. Identify variable starvation level, equivalent to output or speed. This is a primary experimental variable, varied incrementally from one to 16.*

```
  starvation-level
```

*2.4. Set variable energy-saving spatial region behind cells.*

       `draft-zone`

2.5. *Set region of cell attraction. This was held constant at six spaces. The value reflects the tendency for cells to attract to each other.*

       `cohesion-zone`

2.6. *Initialize slug shape. 315° was used for all tests to arrest rapid slug elongation, but a range from lower angles to 315° is reasonable.*

       `in-cone-angle`

2.7. *Permit small random speed reductions.*

       `deceleration-factor`

2.8. *Permit small random accelerations*

       `acceleration-factor`

2.9. *If PCR > 1 and dying cells separate from main group, allow the dying cells to re-attract to the main group, within a limited distance. At one extreme, cells continue to decelerate unless leader pace slows (if parameter is set to zero); at other extreme cells can resume pace of those ahead if within distance behind set by this variable. 1.25 was arbitrarily set so that when cells fall off back, relatively soon they resume the average speed of the group. This is an unnecessary simulation protocol if slug speeds oscillate around the mean threshold speed of the cells (see text, Fig.3), which is the expected actual slug behavior. However, to observe slug dynamics at the larger speed variations of this study, it was useful to include this tunable code.*

       `decel-zone`

3.0. *Set up simulation space. Patch and turtle size may be varied.*

```
to setup
  clear-all
  set-default-shape turtles "circle"
  set-patch-size 2
  resize-world -50 2000 -150 150
  ask n-of peloton-size patches with
   [ pycor > (- 50) and pycor < 60 and pxcor > 300 and pxcor < 400 ]
    [ sprout 1 ]

ask turtles
  [ set size 1
```

3.1. *Simulation runs when PCR values are reported. 100 encompass all possible PCRs contemplated by experiments.*

```
    if pcr <= 100
```

3.2. *Cell migration is left to right, with small random up-down movement.*

```
   [ set heading 90 + random-float 0.02
```

```
    set speed starvation-level
    set color green
    set x-acc 0
    set y-acc 0
```

**4.0.** *Cell-size range from (Bonner and Frascella, 1953, Fig. 1) (120-50) / 120 = 58.3% range. Individual cell sizes are randomly set at simulation start across this range:*

```
  let low 50
  let high 120
  let range high - low + 1
  set MSO (low + random range)]]
  reset-ticks
end
```

**5.0.** *Basic Procedures*

```
to go
  ask turtles  [calculate-position]
  ask turtles  [set-speeds]
  ask turtles  [calculate-separation-factor]
  tick
end
```

**6.0.** *Report cell x-y positions.*

```
to calculate-position
  set x-acc 0
  set y-acc 0
  set neighborhood other turtles in-cone cohesion-zone in-cone-angle
```

**6.1.** *Tabulate and average cell positions.*

```
if any? neighborhood
    [ calculate-average-pos
      calculate-separation-factor ]
end
```

**6.2.** *Small bias to the center, making cells move to the horizontal centre (from Ratamero(2015)).*

```
to put-middle-bias
 let delta 0.005
 if ycor > 0 [ set y-acc (y-acc - delta)]
 if ycor < 0 [ set y-acc y-acc + delta ]
end
```

**7.0.** *Calculate cohesion (modified from Ratamero (2015)). Random movement factors 0.03 and 0.05 reflect speed/direction adjustments that are small relative to cell migration speeds.*

```
to calculate-average-pos

  if pcr < 100
```

```
  [ let x-average mean [ abs xcor ] of neighborhood
    set x-acc x-acc + (x-average - abs xcor) * 0.05
    let y-average mean [ ycor ] of neighborhood
    set y-acc y-acc + (y-average - ycor) * 0.03 ]
end
```

8.0. *Calculate separation force (modified from Ratamero (2015)). Again, random factor 0.25 reflects small speed/direction adjustments relative to cell migration speeds. 2.25 determines collective density, set arbitrarily to reflect reasonable apparent density.*

```
to calculate-separation-factor
  set neighborhood other turtles in-cone 2.25 in-cone-angle
  let x-total sum [ abs xcor / distance myself ] of neighborhood
  let y-total sum [ ycor / distance myself ] of neighborhood
  let x-average sum [ [abs xcor] of myself / distance myself ] of
      neighborhood
  let y-average sum [ [ycor] of myself / distance myself ] of
      neighborhood
  set x-acc x-acc + ((x-average - x-total) * random-float 0.25)
  set y-acc y-acc + ((y-average - y-total) * random-float 0.25)
end
```

9.0. *Calculate the energy-saving (drafting) coefficient (modified from Ratamero (2015)). Draft is nil when > 3 spaces between cells. Check for neighbors in a 90° cone behind each cell; any cells in that region receive full benefit of drafting.*

```
to-report calculate-draft
  let draft 1
  ifelse any? neighborhood
```

9.1. *If more than one cell ahead, drafting quantity is greater than if only one cell ahead. "In-cone" is the angle 90° at the variable distance "draft-zone". "Insiders" are cells within drafting zone.*

```
  [ ifelse count turtles in-cone draft-zone 90 > 1
    [ let neighborhood-draft neighborhood in-cone draft-zone 90
      let insiders [ ]
      if any? neighborhood-draft

    [ foreach sort-by [distance ?1 < distance ?2] neighborhood-draft
      [ let dist distance ?
```

9.2. *Draft coefficient equation from Olds (1998) varies according to spacing such that magnitude diminishes to zero as distance increases to three spaces.*

```
        set draft (draft-coeff - 0.0104 * dist + 0.0452 * dist ^ 2)
        set insiders fput who insiders]]]
```

9.3. *However, if only one cell ahead, give draft in 15-degree cone only. This reflects the fact that drafting effect is smaller when there is only one cell ahead, compared to greater draft when multiple cells fill the area ahead.*

```
    [ let neighborhood-draft neighborhood in-cone draft-zone 15
      let insiders [ ]
```

```
      if any? neighborhood-draft
      [ foreach sort-by [distance ?1 < distance ?2] neighborhood-draft
        [ let dist distance ?
          set draft (draft-coeff - 0.0104 * dist + 0.0452 * dist ^ 2)
          set insiders fput who insiders
          set ycor ycor + y-acc * random-float 0 ]]]]
```

*9.4. If neither draft conditions exist, draft coefficient = 1 means no drafting*

```
      [ set draft 1 ]
        report draft
end
```

*10. Determine cell speeds using fundamental coupling equation (from Trenchard et al. (2015)). PCR is referred to as "CCR" in text and equation (1)*

```
    to set-speeds
      set pcr (power - (power * ( 1 - draft-coefficient ))) / MSO
      set speed starvation-level
      set draft-coefficient calculate-draft
      set neighborhood other turtles in-cone decel-zone in-cone-angle
```

*10.1. Determine "Active" (green) cells. See inequality (4) in text. Again, small random speeds are added to variable leader speed, equivalent to starvation-level.*

```
if MSO > power
   [ set color green
     set speed starvation-level + x-acc + random-normal 0 .001
     set xcor abs xcor + x-acc + random-normal 0 0.001
     set ycor ycor + y-acc + random-normal 0 0.001
     set heading 90 + random-float 0.01
     put-middle-bias ]
```

*10.2. Determine "Suffering" (red) cells. See inequality (3) in text.*

```
if MSO < power and pcr > draft-coeff and pcr < 1
 [  set color red
     set speed starvation-level + x-acc + random-normal 0 .001
     set xcor abs xcor + x-acc + random-normal 0 0.001
     set ycor ycor + y-acc + random-normal 0 0.001
     set heading 90 + random-float 0.01
     put-middle-bias ]
```

*10.21. Alternative code that allows increases in starvation-level for suffering cells randomly between 0 and 1. This permits weaker but more active cells to migrate to the anterior.*

```
 [ set color red
       set speed starvation-level + x-acc + random-float 1
       set xcor abs xcor + x-acc + random-float 1 ]
```

**10.3.** *Determine "Dying or dead" (yellow) cells. See inequality (2) in text.*

```
 if pcr >= 1
  [ set color yellow
    set ycor ycor + y-acc + random-normal 0 0.001
    put-middle-bias ]
```

**11.0.** *Deceleration algorithm allows cells to decelerate if over threshold. See Part 12 below for equations that determine deceleration magnitude, from Trenchard et al.(2015).*

**11.1.** *Reduce speed if PCR > 1*

```
 ifelse pcr > 1 and any? neighborhood
   [ set slowing-speed speed - (V_reduction + random-float deceleration-
       factor)
     set xcor (abs xcor - (V_reduction + random-float deceleration-factor))
     put-middle-bias ]
```

**11.2.** *Otherwise migrate at regular speed if PCR is <= 1.*

```
 [ set speed starvation-level + x-acc + random-float acceleration-factor
   set xcor abs xcor + x-acc + random-float acceleration-factor
   put-middle-bias ]
  end
```

**12.0.** *Power output is determined by accounting for a set of friction/drag factors. Here, human-scale factors are retained (see text for explanation).*

```
to-report power
    set speed starvation-level + x-acc + random-float acceleration-factor
    let A 0.639      Frontal area in m^2 (area of typical cyclist)
    let Cw 0.5       Drag coefficient
    let cm 0.015     Coefficient for power losses due slippage
    let Rho 1.226    Air density kg m^3
    let Crr .004     Coefficient of rolling resistance
    let wkg 75       Combined weight of cyclist and bicycle
    let fw 0.5 * cw * A * Rho * ((speed + wind-speed*) ^ 2)  Here wind-speed
                                                               is 0
    let Crv 0.1      Coefficient for velocity-dependent dynamic rolling
                  resistance, here approximated with 0.1
    let Crvn Crv * (cos gradient) Coefficient for the dynamic rolling
                          resistance, normalized to road inclination;
                        CrVn = CrV*cos(ß)
    let frl Wkg * 9.8 * Crr
    let Frg 9.8 * Wkg * ((crr * cos gradient) + (sin gradient))
    let fsl Wkg * 9.8 * gradient
    let power-output (fw + fsl + frl) * (speed)
    report power-output
```

**12.1.** *If a cell is over MSO threshold, determine deceleration magnitude to bring cell back under threshold. First find power-output of leader.*

```
to-report V_eff
```

```
let Peff (MSO * PCR) / calculate-draft
```

12.2. *Next convert power-output of leader to speed.*

```
    let A 0.639
    let Wkg 75
    let Cw 0.5
    let cm 0.015
    let Rho 1.226
    let Crr .004
    let Crv 0.1
    let Crvn Crv * (cos gradient)
    let frl Wkg * 9.8 * Crr
    let Frg 9.8 * Wkg * ((crr * cos gradient) + (sin gradient))
    let fsl Wkg * 9.8 * gradient
    let a1 0.5 * A * cw * Rho
```

12.2. *Use the following equations for speed-to-power conversion. See Trenchard et al. (2015) for details and references.*

```
    let b 2 * a1 * wind-speed*;
    let c 9.8 * Wkg * (crr + gradient)
    let d Peff * (-1)
    let f (3 * c / a1 - b ^ 2 / a1 ^ 2) / 3
    let g ( (2 * (b ^ 3) / (a1 ^ 3)) - (9 * b * (c / a1 ^ 2)) + (27 * (d
        /a1)))/ 27
    let h ((g ^ 2) / 4) + ((f ^ 3) / 27)
    let r  (-(g / 2) + (h ^ (1 / 2)))
    let s r ^ (1 / 3)
    let t (-(g / 2) - (h ^ (1 / 2)))
    let u 0
    ifelse t > = 0 [set u  (t ^ (1 / 3))]
    [set u (-(- t ^ (1 / 3)))]
    let PeffV s + u - (b / 3 * a1)

report PeffV
```

12.3. *Then find the threshold power-output of follower in terms of power when PCR = 1. See Trenchard et al. (2015) for further explanation.*

```
to-report P1V

set P1 (MSO / calculate-draft)

    let A 0.639
    let Cw 0.5
    let cm 0.015
    let Rho 1.226
    let Crr .004
    let Crv 0.1
    let Crvn Crv * (cos gradient)
    let frl Wkg * 9.8 * Crr
    let Frg 9.8 * Wkg * ((crr * cos gradient) + (sin gradient))
    let fsl Wkg * 9.8 * gradient
```

```
        let a1 0.5 * A * cw * Rho
        let c 9.8 * Wkg * (crr + gradient)
```

12.4. *Again, use the following equations for speed-to-power conversion.*

```
        let b 2 * a1 * wind-speed*
        let d P1 * (-1)
        let f (3 * c / a1 - b ^ 2 / a1 ^ 2) / 3
        let g ( (2 * (b ^ 3) / (a1 ^ 3)) - (9 * b * (c / a1 ^ 2)) + (27 * (d / a1))) / 27
        let h ((g ^ 2) / 4) + ((f ^ 3) / 27)
        let r  (-(g / 2) + h ^ (1 / 2))
        let s r ^ (1 / 3)
        let t (-(g / 2) - h ^ (1 / 2))
        let u 0
        ifelse t >= 0 [set u  (t ^ (1 / 3))]
        [set u (-(- t ^ (1 / 3)))]
        let PV s + u - (b / 3 * a1)
report Pv
end
```

12.5. *Find deceleration magnitude required for cell over MSO threshold to reduce speed to be under threshold, in terms of speed (c.f. Section 11.1.)*

```
to-report V_reduction
  set p_reduction (V_eff - P1V)
  report p_reduction
end
```